\documentclass[pre,nofootinbib,superscriptaddress,onecolumn,preprintnumbers]{revtex4}
\usepackage{graphicx}
\usepackage{amsmath,amssymb}
\usepackage{color}

\renewcommand{\eqref}[1]{Eq.(\ref{#1})}
\newcommand{\eqrefs}[1]{Eqs.(\ref{#1})}

\begin{document}
\title{Modified gravity in the presence of matter creation: scenario for the late Universe}

\author{G. Montani}
\email{giovanni.montani@enea.it}
\affiliation{ENEA, Nuclear Department, C. R. Frascati, Via E. Fermi 45,  Frascati, 00044 Roma, Italy}
\affiliation{Physics Department, ``Sapienza'' University of Rome,  P.le Aldo Moro 5, 00185 Roma, Italy}

\author{N. Carlevaro}
\email{nakia.carlevaro@enea.it}
\affiliation{ENEA, Nuclear Department, C. R. Frascati, Via E. Fermi 45,  Frascati, 00044 Roma, Italy}

\author{M. De Angelis}
\email{mdeangelis1@sheffield.ac.uk}
\affiliation{School of Mathematics and Statistics, University of Sheffield, Hounsfield Road, Sheffield S3 7RH, UK}

\begin{abstract}
We consider a dynamic scenario for characterizing the late Universe evolution, aiming to mitigate the Hubble tension. Specifically, we consider a metric $f(R)$ gravity in the Jordan frame which is implemented to the dynamics of a flat isotropic Universe. This cosmological model incorporates a matter creation process, due to the time variation of the cosmological gravitational field. We model particle creation by representing the isotropic Universe (specifically, a given fiducial volume) as an open thermodynamic system. The resulting dynamical model involves four unknowns: the Hubble parameter, the non-minimally coupled scalar field, its potential, and the energy density of the matter component. We impose suitable conditions to derive a closed system for these functions of the redshift. In this model, the vacuum energy density of the present Universe is determined by the scalar field potential, in line with the modified gravity scenario. Hence, we construct a viable model, determining the form of the $f(R)$ theory a posteriori and appropriately constraining the phenomenological parameters of the matter creation process to eliminate tachyon modes. Finally, by analyzing the allowed parameter space, we demonstrate that the Planck evolution of the Hubble parameter can be reconciled with the late Universe dynamics, thus alleviating the Hubble tension.
\end{abstract}

\maketitle

\section{Introduction}
Over the past decade, cosmological studies of the late Universe have uncovered a significant discrepancy in data, known as the ``Hubble tension''. This tension arises from the differing values of the Hubble constant ($H_0$) measured by the SH0ES Collaboration using Type Ia Supernova (SNIa) data \cite{Scolnic_2022,2018ApJ...859..101S,Brout:2022vxf,riess2022apjl} and those obtained by the Planck Satellite Collaboration \cite{Planck:2018vyg}. The discrepancy, approximately $5\sigma$, presents a perplexing challenge, prompting new considerations regarding the dynamics of the late Universe. For a series of discussions on this topic, see \cite{schiavone_mnras,deangelis-fr-mnras,2024PDU....4401486M,2024PhRvD.109b3527A,Nojiri:2022ski,Odintsov:2020qzd,2016RPPh...79i6901W,naidoo2024PhRvD,DAINOTTI202430,Dainotti2023ApJ...951...63D}. {We also underline how, nonetheless, in the presence of such a tension, the most commonly accepted cosmological model which describes the evolution of the Hubble parameter is the so-called Lambda Cold Dark Matter ($\Lambda$CDM) model \cite{bib:weinberg-2008,DiValentino:2021izs}.This corresponds to including, in the Universe dynamics, a matter-like contribution and a cosmological constant.} The possibility of interpreting the Hubble tension as a continuous effective variation of the Hubble constant based on redshift---where its value appears to depend on the redshift region of the astrophysical sources used for its determination---has been supported by analyses in \cite{apj-powerlaw,galaxies10010024,2024MNRAS.530.5091X}, and is also discussed in \cite{Krishnan:2020vaf,kazantzidis}. Moreover, the coexistence of SNIa and baryon acoustic oscillations (BAOs) \cite{2021MNRAS.505.3866E} (and references therein) data within the same redshift region, and their resulting tension, as BAO provides a $H_0$ value compatible with that from Planck, has led to the development of early dark energy (DE) models \cite{giare2024arXiv240412779G} and a suitable combination of early and late modified dynamics \cite{2023Univ....9..393V,2020PhRvD.102b3518V}. In particular, the analysis in \cite{deangelis-fr-mnras} proposed a specific $f(R)$ model, examined within the so-called Jordan frame \cite{Sotiriou-Faraoni:2010,NOJIRI201159}, to effectively address the Hubble tension in the presence of evolutionary DE, which transits to a phantom contribution at $z>1$. This model presents an intriguing scenario where the Hubble tension is essentially resolved at $z\gtrsim 2$, since the non-minimally coupled scalar field shows a monotonically increasing behaviour toward an asymptote. This asymptotic configuration aligns with the Planck value for the Hubble constant, occurring at a relatively low redshift.


{In this work, we explore a similar approach, but rather than initially assuming the existence of evolutionary dark energy (DE), we consider a more natural physical scenario. This scenario involves the gravitational field generating weakly massive particles due to its time variation, effectively introducing a radiation component in the late Universe. Specifically, the matter creation process is treated phenomenologically, by regarding the Universe as an open thermodynamic system, as discussed in previous works like \cite{2001CQGra..18..193M,1992PhLA..162..223C}. The particle creation rate is determined using an \emph{ansatz} in the form of a power-law in the Hubble parameter. For a more comprehensive understanding of the matter creation process across the Universe, please see \cite{erdem24a,2015PhRvD..91f3526N,2016GReGr..48..107N,2014JCAP...10..042L}.

After constructing the dynamical system governing the late Universe dynamics and incorporating the creation of a radiation component by the gravitational background, we demonstrate how the Hubble tension could be effectively alleviated within certain favourable regions of the parameter space. This framework offers a promising depiction of the underlying mechanism potentially resolving such a cosmological issue. It is noteworthy that the current-day radiation created does not exceed a few percent, and the anticipated weakly interacting nature of these particles explains why this generated energy density remains indirectly observed in the actual Universe.}

The paper is structured as follows. In Section \ref{sec2}, a phenomenological approach to the matter creation process is presented and the dynamics for the created energy density is fixed. In Section \ref{sec3}, the formulation of a metric $f(R)$ gravity in the Jordan frame is reviewed and the basic features of the proposed model are traced. In Section \ref{sec4}, the dynamical model corresponding to the evolution of a flat isotropic Universe in the considered modified scenario is constructed. The set of free parameters is described and the initial conditions for constructing the numerical solutions are given. In Section \ref{sec5}, the free parameter space is numerically investigated generating a triangular plot. A privileged set of parameters is then identified and the profile of the Hubble parameter is provided in order to show the capability of the model to alleviate the Hubble tension.

\section{Matter Creation Process}\label{sec2}
In the thermodynamic framework presented in this work, the concept of matter creation from a time-varying gravitational field relies on a simple phenomenological representation. {Let us start from the first principle, as follows:
\begin{equation}
dU = -pdV + \delta Q + \mu dN\,,
\label{mcm1}
\end{equation}
which has to be combined with the heat change $\delta Q$, expressed by means of the second thermodynamics principle as $\delta Q=TdS$}. Here, $U$ is the internal energy, $p$ is the pressure, $T$ is the temperature, $S$ is the entropy, $\mu$ is the chemical potential, $N$ is the particle number, and $V$ is the volume of the considered system. Introducing now the expressions $U=\rho V$, $S = \sigma N$, and $\mu = (\rho + p )V/N - \sigma T$ (in which $\rho$ is the {total} energy density of the system and $\sigma$ is the entropy per particle), we can rewrite Equation (\ref{mcm1}) as follows:
\begin{equation}
	d\rho = - \left(\rho + p\right)\left( 1 - \frac{d\ln N}{d\ln V}\right) \frac{dV}{V} + T\frac{N}{V}d\sigma\,.
	\label{mcm2}
\end{equation}
The difference from an iso-entropic system is that we only need to ensure the preservation of entropy per particle by setting $d\sigma = 0$. Hence, it is clear that, since $\sigma = S/N$, 
matter creation results in an increase in the entropy of the system. Therefore, Equation (\ref{mcm2}) simplifies as follows: 
\begin{equation}
	\frac{d\rho}{d\tau} = 
	- \left( \rho + p\right) \left( 1 - \frac{d\ln N}{d\ln V}\right) \frac{d\ln V}{d\tau}\,,
	\label{mcm3}
\end{equation}
once it is applied to a homogeneous background and its evolution is tracked with respect to a clock labelled $\tau$. {Here, the time variable $\tau$ thus denotes the specific time variable associated with the considered dynamical system and, in what follows, it will be identified with the cosmological Gaussian time.} The line element of a flat isotropic Universe \cite{2020MNRAS.496L..91E} reads as follows:
\begin{equation}
ds^2 = -dt^2 + a^2(t)	\left( dx^2 + dy^2 + dz^2\right)\, , 
\label{mcm4}
\end{equation}
in which $t$ denotes the synchronous time ($c=1$), and $\{x,y,z\}$ are the Cartesian coordinates. Moreover, $a(t)$ stands for the cosmic scale factor responsible for the Universe expansion. 

In terms of the Hubble parameter $H(t)\equiv \dot{a}/a$ (the dot denoting synchronous time differentiation), a reliable \emph{ansatz} for the matter creation is as follows:
\begin{equation}
	\frac{d\ln N}{d\ln V} = 
	\left( H/\bar{H}\right)^{\beta}\,, 
	\label{ans}
\end{equation}
where $\beta$ and $\bar{H}$ are positive free parameters of the model and the considered fiducial volume can be set as $V=a(t)^3V_0$, with $V_0$ as a generic coordinate volume which does not enter the dynamics. 

{Since the present-day Universe expansion rate is rather slow, it is natural to argue that the gravitational field is generating very low-mass particles \cite{birrelldavies,WALDbook}.} Given that, we address this process to a radiation-like component energy density $\rho_r$. 
According to the analysis above, the dynamics of such an emerging radiation contribution is governed by the following equation {(i.e., we use the equation of state $p=w \rho$ with $w=1/3$):}
\begin{equation}
	\dot{\rho}_r = -4H\rho_r\left( 1- \left(H/\bar{H}\right)^{\beta}\right)\,.
	\label{baeq}
\end{equation}
{This equation comes from Equation (\ref{mcm3}), once we use $\tau=t$, the \textit{ansatz} in Equation (\ref{ans}), and by implementing $d\ln \!V /dt = 3H$.} In the following sections, we embed this mechanism in the context of a metric $f(R)$ gravity in the Jordan frame.

\section{Modified Cosmological Dynamics in the Jordan Frame}\label{sec3}
In the Jordan frame, the action of a metric $f(R)$ gravity can be cast as follows \cite{Sotiriou-Faraoni:2010}:
\begin{equation}
	\mathcal{S} = \frac{1}{2\chi}\int d^4x\sqrt{-g}\left(\phi R - V(\phi )\right)
	\, , 
	\label{ent1}
\end{equation}
where $\chi$ denotes the Einstein constant, $g$ and $R$ are the metric determinant and the Ricci scalar, respectively, while the non-minimally coupled scalar field $\phi$ is the independent degree of freedom expressing the modified gravity formulation. In particular, the potential term $V(\phi)$ is linked to the specific form of the function $f(R)$ via the following relation:
\begin{equation}
	f(R(\phi )) = \phi \frac{dV}{d\phi} - V(\phi )
	\, , 
	\label{ent2}
\end{equation}
which comes from the basic definition $\phi \equiv df/dR$ and from the substitution of the field equation into the expression of $V$, obtained by varying the action \eqref{ent1} with respect to $\phi$, i.e., $R = dV/d\phi$.
A basic viability condition for the choice of a given $f(R)$ model is that the scalar mode possesses a positively defined quadratic mass, according to the following definition \cite{Olmo:2005hc}: 
\begin{equation}
	\mu^2_{\phi}\equiv \frac{1}{3}\left( \phi \frac{d^2V}{d\phi ^2} - \frac{dV}{d\phi}\right) \ge 0
	\, .
	\label{ent3}
\end{equation}
The variation in the action (\ref{ent1}) with respect to the metric tensor yields the vacuum Einstein equations of the extended scalar--tensor theory. By taking the trace of these equations and incorporating the condition $R = dV/d\phi$, we can derive a Klein--Gordon-like equation for the scalar field. Introducing a matter source is straightforward; it involves a conserved energy--momentum tensor that represents the specific physical system under consideration. The trace of this tensor also contributes to the Klein--Gordon-like equation for the scalar field, as discussed by \cite{Sotiriou-Faraoni:2010}.

We now specify the field equation for the line element of a flat isotropic Universe (see \eqref{mcm4}) in the presence of a co-moving total energy density $\rho_{tot}$. In particular, we consider the $00$-component of the Einstein Equation \cite{bib:weinberg-2008}, as follows:
\begin{equation}
	H^2 = \frac{\chi \rho_{tot}}{3\phi} - H\frac{\dot{\phi}}{\phi} + \frac{V(\phi )}{6\phi}
	\, .
	\label{ent4}
\end{equation}
Here, $\rho_{tot} = \rho_m + \rho_r$, 
where $\rho_m$ denotes the (dark and baryonic) matter component of the Universe, verifying the standard conservation law {i.e., the divergenceless nature of the corresponding energy--momentum tensor, as follows:}
\begin{equation}
	\dot{\rho}_m + 3H\rho_m = 0 \, \rightarrow \, \rho_m = \rho_m^0(1 + z)^3
	\, , 
	\label{ent5}
\end{equation}
in which $\rho_m^0$ denotes the present-day value of $\rho_m$. We introduced the redshift variable $z(t) = 1/a(t) - 1$ (we set the present value of the cosmic scale factor equal to unity). {It is worth noting that the evolution of the Universe matter component is not affected by the matter creation process. Actually, the radiation creation, as described in Equation (\ref{baeq}), does not follow the divergenceless character of a prefect fluid with $w=1/3$. However, it corresponds to a perfect fluid with the following time-varying equation of a state parameter:
\begin{equation}
w_r(t) = \frac{1}{3}- \frac{4}{3}\left(
\frac{H}{\bar{H}}\right)^{\beta}
\, .
\label{cl}
\end{equation}
It is the conservation law of a such an energy--momentum tensor which provides \mbox{Equation~(\ref{baeq}).}} 
The second basic equation of the cosmological dynamics reads as follows:
\begin{equation}
	\frac{dV}{d\phi} = R = 12H^2 + 6\dot{H}
	\, .
	\label{ent6}
\end{equation}
This approach is particularly suitable for determining a posteriori the form of the $f(R)$ gravity that can help alleviate the Hubble tension.

\section{Reduced Dynamics}\label{sec4}
In order to transform the potential $V(\phi)$ from an ingredient which is assigned via the function $f(R)$ into a dynamical variable $V(t)$, we impose on \eqref{ent4} the following two conditions: 
\begin{align}
V(\phi ) = 2\chi\rho_{\Lambda} + G(\phi )\,,\qquad
6H\dot{\phi} = G\, , 
\label{ent7}
\end{align}
where $\rho_{\Lambda}$ is the constant value of the Universe energy density and $G$ is a generic functional form, to be dynamically determined. Clearly, these two relations play a crucial role in giving Equation (\ref{ent4}) a form that resembles the dynamics of a $\Lambda$CDM model for the Universe, {as specified at the beginning of the next section}. Introducing the critical parameters in place of the energy density according to the relations $\Omega_m^0 = \chi \rho_m/3H_0^2$, $\Omega_\Lambda^0 = \chi \rho_\Lambda/3H_0^2$, and $\Omega_r = \chi \rho_r/3H_0^2$, where $H_0$ is the present day value of the Hubble constant, we can rewrite Equation~(\ref{ent4}) into the following form:
{\begin{equation}
H^2(z) = \frac{H_0^2}{\phi(z)}\left( \Omega_m^0 (1+z)^{3} + \Omega_{\Lambda}^0+\Omega_r(z)\right)
\, , \label{ent8}
\end{equation}}
where $\Omega_r^0=\Omega_r(z=0)$ and the relation $\Omega_{\Lambda}^0=1-\Omega_m^0-\Omega_r^0$ must be satisfied, given that we set $\phi(z=0)=1$. This equation is coupled with the dynamical system of Equations (\ref{baeq}), (\ref{ent6}), and (\ref{ent7}) recast as follows (the prime denotes $z$-differentiation): 
\begin{align}
	\phi^{\prime} &= - \frac{G(z)}{6(1+z)H^2}
	\, ,
	\label{ent9}\\
	G^{\prime} &= 3\phi ^{\prime}\left(4H^2 - 2(1+z)HH'\right)
	\, ,
	\label{ent10}\\
\Omega_r' &= 4(1+z) \Omega_r\left( 1 - \left(H/\bar{H}\right)^{\beta}\right)
	\, .
	\label{ent11}
\end{align}
The ratio of \eqref{ent9} and \eqref{ent10} gives us the following relation \cite{deangelis-fr-mnras}:
\begin{equation}
	G(z) = - 6A^2 H(z)(1+z)^{-2}
	\, , 
	\label{ent12}
\end{equation}
where $A^2$ is a positive integration constant. Substituting the expression above into \mbox{\eqref{ent9}}, we obtain the following:
\begin{equation}
	\phi^{\prime} = \frac{A^2}{H(z)(1+z)^{3}}
	\, ,
	\label{ent13}
\end{equation}
which describes the increasing behaviour of $\phi$ with the time variable $z$. 

In this scheme, the dynamical system thus reduces to the following: 
\begin{align}
	&\phi^{\prime} = A_0^2\;(1+z)^{-3} \;\big(\phi(z)^{-1}(\Omega_m^0 (1+z)^{3} + 1-\Omega_m^0-\Omega_r^0 + \Omega_r(z) )\big)^{-1/2}
	\, , \label{fs1}\\
	 &\Omega _r' = 4(1+z)^{-1} \Omega _r(x) \Big( 1 - \big(\bar{H}_0^{-2}\phi(z)^{-1}(\Omega_m^0 (1+z)^{3} + 1-\Omega_m^0-\Omega_r^0 + \Omega_r(z) )\big)^{\beta/2}\Big)
	\, , \label{fs2}
\end{align}
where $A^2_0 \equiv A^2/H_0$ and $\bar{H}_0\equiv \bar{H}/H_0$. Once we have calculated the function $\phi(z)$, we can also determine the potential $G(\phi)$ and, finally, the form of the resulting $f(R)$ gravity.

\section{Numerical Analysis}\label{sec5}
In this section, we provide a comprehensive description of the methodology used for integrating the model. Specifically, we numerically derive the dynamical forms of $\phi(z)$ and $\Omega_r(z)$, which ultimately lead to the final expression of $H(z)$ in \eqref{ent8}. {To compare our results with the standard two flat $\Lambda$CDM forms of the Hubble parameter constrained by the early Universe data \cite{Planck:2018vyg} and by the Phanteon+ dataset \cite{Brout:2022vxf}, we introduce the following quantities:}
\begin{align}
&H_{Pl}(z)=H_0^{Pl}(\Omega_m^{0Pl} (1+z)^{3}+(1-\Omega_m^{0Pl}))^{1/2}\;,\label{hplanck}\\
&H_{SN}(z)=H_0^{SN}(\Omega_m^{0SN} (1+z)^{3}+(1-\Omega_m^{0SN}))^{1/2}\;,\label{hsn}
\end{align}
with ($H$ and $H_0$ are in units of km s$^{-1}$ Mpc$^{-1}$):
\begin{align}
&H_0^{Pl}=67.4\pm0.5 \;,\,\,\qquad \Omega_m^{0Pl}=0.315\pm0.007\;,\\
&H_0^{SN}=73.6\pm1.1 \;,\qquad \Omega_m^{0SN}=0.334\pm0.018\;,
\end{align}
In order to study the viability of the addressed scenario and also its capability in alleviating the Hubble tension, we explore the full free parameter space of the model $\{H_0,\,\Omega_m^0,\,\Omega_r^0,\,\bar{H}_0,\,A_0,\,\beta\}$. Equations (\ref{fs1}) and (\ref{fs2}) are numerically solved, spanning a grid of 15 values for each parameter, and thus collecting $15^6$ sampled different models. The ranges are defined according to previously conducted tests on the integrability of the system and phenomenological considerations: we have set $72.5 \leq H_0\leq 74.7$, $0.25 \leq \Omega_m^0\leq 0.335$, $0.01 \leq \Omega_r^0\leq 0.15$, $0.5 \leq \bar{H}_0\leq 1.5$, $0.1 \leq A_0\leq 0.5$, and $0.5 \leq \beta\leq 1.5$. We note that the selected values of $H_0$ are in accordance with the measurements of the Pantheon+ analysis \cite{Brout:2022vxf}, while $\Omega_r^0$ is assumed, as already stated, to remain a small contribution to the energy density of the Universe.

\begin{figure}
\hspace{-25pt}\includegraphics[width=14cm]{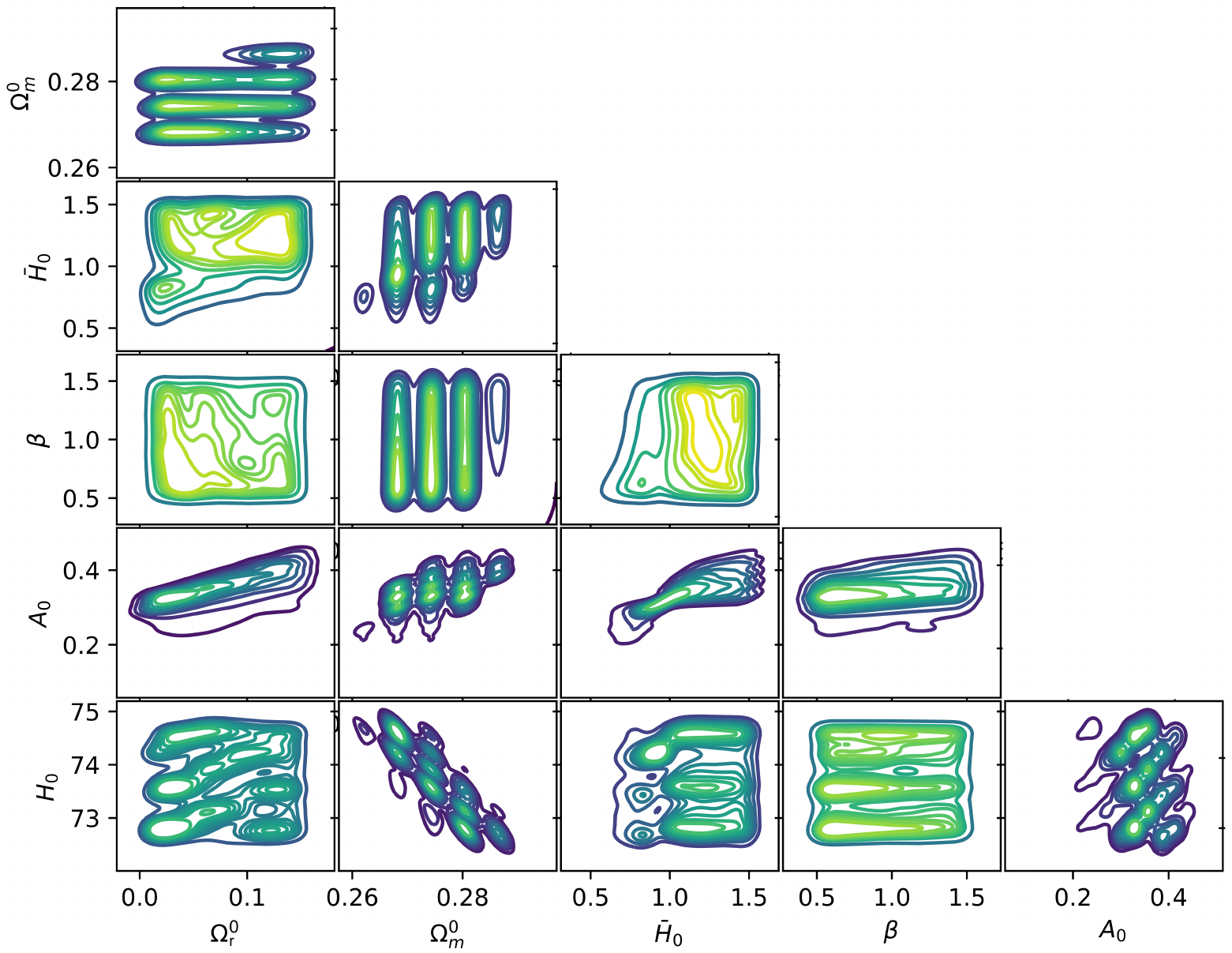}
\caption{{Density} plot for all the model’s parameters. The yellow region indicates the most frequent values preferred by the model.}
\label{figplot}
\end{figure}

The obtained models are then filtered requiring non-tachyonic modes, i.e., that \mbox{\eqref{ent3}} is guaranteed, and we further impose the condition that the normalized (by $H_0^2$) squared mass is less or equal to unity, today. This point is relevant in order for the considered $f(R)$ model to obey the so-called \lq\lq chameleon\rq\rq \, mechanism \cite{Brax:2008hh}. The resulting models are thus physically viable and, to study the efficiency in order to alleviate the tension, we also require that $0.999<H/H_{Pl}|_{z=10}<1.001$. With this procedure, we finally obtain around $5 \times 10^3$ sampled models. We then convolve the data with a normal distribution to create a smooth density estimate. A kernel density estimate plot is a visualization method used to depict the distribution of observations in a dataset, akin to a histogram. The results are depicted in Figure \ref{figplot}. 

\begin{figure}
\includegraphics[width=9.5cm]{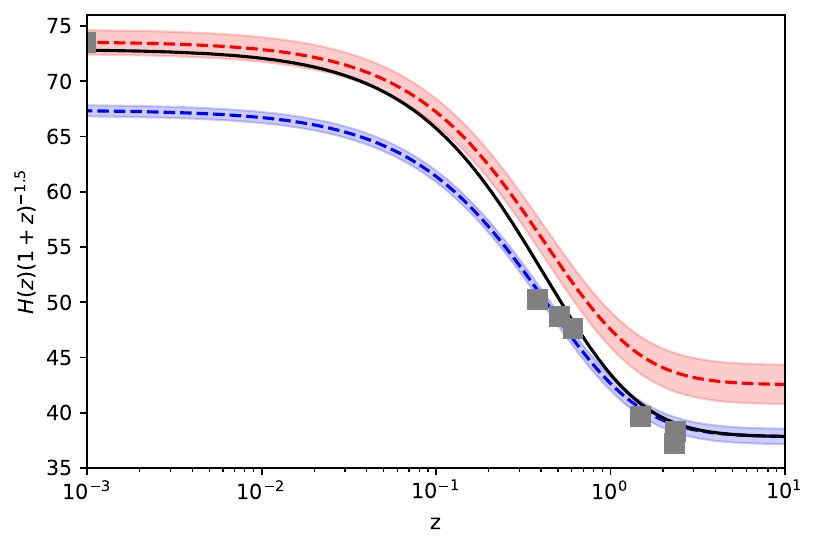}
\caption{Plot of $H(z)$ (black) using the parameters in \eqref{parset} and the profiles (with the corresponding errors) of $H_{Pl}(z)$ (blue) and $H_{SN}(z)$ (red). {Grey squares represent the SH0ES prior and six relevant measurements for BAO sources for $0.3<z<3$.}}
\label{figH}
\end{figure}

\begin{figure}
\includegraphics[width=9.5cm]{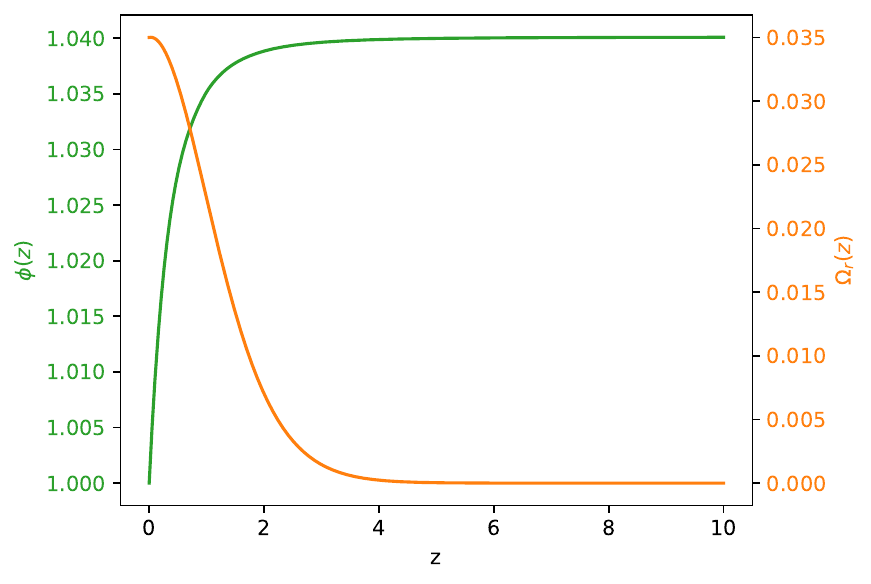}
\caption{Plot of $\phi(z)$ (green) and $\Omega_r(z)$ (orange) integrated from \eqrefs{fs1} and (\ref{fs2}), implementing the parameters in \eqref{parset}.}
\label{figOphi}
\end{figure}

From this analysis, preferred regions can be individualized providing the most frequent parameter sets. As an example of the model's capability in alleviating the Hubble tension, we select the following:
\begin{align}\label{parset}
H_0=72.9\,,\quad
\Omega_m^0=0.28\,,\quad
\Omega_r^0=0.035\,,\quad
\bar{H}_0=1.01\,,\quad
A_0=0.33\,,\quad
\beta=0.7\,.
\end{align}
In Figure \ref{figH}, we plot the resulting evolution of $H(z)$ from \eqref{ent8}, with this parameter setup, together with the curves in \eqref{hplanck} (blue) and \eqref{hsn} (red), considering the corresponding errors. The tension alleviation emerges clearly, and the $H(z)$ profile overlaps $H_{Pl}(z)$ already at $z\simeq3$, but reaches higher values of the Hubble constant in $z=0$. {In the figure, we also depict six relevant measurements for BAO sources (in the range $0.3<z<3$) \cite{2017MNRAS.470.2617A,2019A&A...629A..86B,2019A&A...629A..85D,2021MNRAS.500.1201H} and the SH0ES prior for $H(z)$ today. This clearly indicates the capability of the addressed model to also alleviate the tension derived from different late Universe sources.} In Figure \ref{figOphi}, we instead plot the functions $\phi$ and $\Omega_r$ as functions of $z$. The curves are obtained by integrating Equations (\ref{fs1}) and (\ref{fs2}), assuming the choice of the parameters as in \eqref{parset}. As expected, for $z>3$, $\Omega_r(z)$ approaches zero, while $\phi(z)$ is frozen as a plateau. 

\unskip

This study of the parameter space is essential for guiding the real data analysis procedure, clearly indicating the model's viability for interpreting the Hubble tension through the missing physics that must be added to the $\Lambda$CDM formulation to reconcile it with observations. The numerical study highlights how the additional components of modified metric $f(R)$ gravity and the radiation term generated by the cosmological background are crucial for mitigating the Hubble tension within the framework of a physical theory. In this context, a key aspect of the numerical filtering procedure is ensuring that the modified gravity is not associated with a tachyonic mode.

\section{Conclusions}
We constructed a revised dynamical model for the late Universe based on both a metric $f(R)$ gravity in the Jordan frame, similar to the approach in \cite{deangelis-fr-mnras}, and a phenomenological model for matter creation associated with the time variation of the cosmological gravitational field. The matter creation is described by treating the Universe as an open thermodynamic system, with the rate of particle creation phenomenologically regulated by a power-law of the Hubble parameter. The form of the $f(R)$ gravity is not assigned a priori. Indeed, by imposing suitable conditions on the generalized Friedmann equation, the potential of the non-minimally coupled scalar field $V(z)\equiv V(\phi (z))$ becomes a dynamical variable, along with $\phi(z)$ and $\Omega_r(z)$. Consequently, it is possible to reconstruct the form of $V(\phi)$ and thus determine the function $f(R)$ governing the modified gravity.

{We developed a dynamical model with six free parameters, $\{H_0,\,\Omega_m^0,\,\Omega_r^0,\,\bar{H}_0,\,A_0,\,\beta\}$, aimed at alleviating the Hubble tension. Our numerical analysis involved a thorough screening of all possible solutions, retaining only those parameter sets that ensured the physical consistency of the proposed scenario. These solutions also had to meet two criteria: the resulting Hubble parameter needed to exhibit the required asymptotic $\Lambda$CDM behaviour (achieved at low $z$ values) and allow for a Hubble constant compatible with the SH0ES collaboration observations. Identifying a preferred set of parameters provided significant insights into how real data comparisons could impact the parameter space. This approach could successfully address the Hubble tension, as the Hubble parameter rapidly converges to the $\Lambda$CDM model with Planck-measured parameters as the redshift increases. }

We conclude by emphasizing that the privileged values of $\Omega_r^0$ were found to be very small, making this component reliably unobserved through direct measurement. The particles forming this radiation component, such as sterile neutrinos \cite{2017PhR...711....1A}, very weakly interact, which explains their elusive nature. The key takeaway from this study is that the Hubble tension can be effectively addressed by combining metric $f(R)$ gravity with additional modifications to the standard $\Lambda$CDM model.

\acknowledgments{MDA thanks William Giarè for valuable discussions.}


\begin{thebibliography}{10}
\ProvideTextCommand{\guillemotleft}{OT1}{%
  \leavevmode\raise .27ex\hbox{$\scriptscriptstyle\ll$}}
\ProvideTextCommand{\guillemotright}{OT1}{%
  \leavevmode\raise .27ex\hbox{$\scriptscriptstyle\gg$}}
\newcommand{\enquote}[1]{\guillemotleft#1\guillemotright}

\bibitem{Scolnic_2022}
Dan~M. Scolnic, Dillon Brout, Anthony Carr, Adam~G. Riess, Tamara~M. Davis, Arianna Dwomoh, David~O. Jones, Noor Ali, Pranav Charvu, Rebecca Chen, Erik~R. Peterson, Brodie Popovic, Benjamin~M. Rose, Charlotte~M. Wood, Peter~J. Brown, Ken Chambers, David~A. Coulter, Kyle~G. Dettman, Georgios Dimitriadis, Alexei~V. Filippenko, Ryan~J. Foley, Saurabh~W. Jha, Charles~D. Kilpatrick, Robert~P. Kirshner, Yen-Chen Pan, Armin Rest, Cesar Rojas-Bravo, Matthew~R. Siebert, Benjamin~E. Stahl and WeiKang Zheng, \emph{ApJ} \textbf{938}, 113 (2022).

\bibitem{2018ApJ...859..101S}
D.~M. {Scolnic}, D.~O. {Jones}, A.~{Rest}, Y.~C. {Pan}, R.~{Chornock}, R.~J. {Foley}, M.~E. {Huber}, R.~{Kessler}, G.~{Narayan}, A.~G. {Riess}, S.~{Rodney}, E.~{Berger}, D.~J. {Brout}, P.~J. {Challis}, M.~{Drout}, D.~{Finkbeiner}, R.~{Lunnan}, R.~P. {Kirshner}, N.~E. {Sanders}, E.~{Schlafly}, S.~{Smartt}, C.~W. {Stubbs}, J.~{Tonry}, W.~M. {Wood-Vasey}, M.~{Foley}, J.~{Hand}, E.~{Johnson}, W.~S. {Burgett}, K.~C. {Chambers}, P.~W. {Draper}, K.~W. {Hodapp}, N.~{Kaiser}, R.~P. {Kudritzki}, E.~A. {Magnier}, N.~{Metcalfe}, F.~{Bresolin}, E.~{Gall}, R.~{Kotak}, M.~{McCrum} and K.~W. {Smith}, \emph{ApJ} \textbf{859}, 101 (2018).

\bibitem{Brout:2022vxf}
Dillon Brout \emph{et~al.}, \emph{ApJ} \textbf{938}, 110 (2022).

\bibitem{riess2022apjl}
Adam~G. {Riess}, Wenlong {Yuan}, Lucas~M. {Macri}, Dan {Scolnic}, Dillon {Brout}, Stefano {Casertano}, David~O. {Jones}, Yukei {Murakami}, Gagandeep~S. {Anand}, Louise {Breuval}, Thomas~G. {Brink}, Alexei~V. {Filippenko}, Samantha {Hoffmann}, Saurabh~W. {Jha}, W.~{D'arcy Kenworthy}, John {Mackenty}, Benjamin~E. {Stahl} and WeiKang {Zheng}, \emph{ApJ Lett.} \textbf{934}, L7 (2022).

\bibitem{Planck:2018vyg}
N.~Aghanim \emph{et~al.}, \emph{A\& A} \textbf{641}, A6 (2020), [Erratum: Astron.Astrophys. 652, C4 (2021)].

\bibitem{schiavone_mnras}
Tiziano Schiavone, Giovanni Montani and Flavio Bombacigno, \emph{Mon. Not. RAS} \textbf{522}, L72 (2023).

\bibitem{deangelis-fr-mnras}
Giovanni {Montani}, Mariaveronica {De Angelis}, Flavio {Bombacigno} and Nakia {Carlevaro}, \emph{Mon. Not. RAS} \textbf{527}, L156–L161 (2024).

\bibitem{2024PDU....4401486M}
Giovanni {Montani}, Nakia {Carlevaro} and Maria~Giovanna {Dainotti}, \emph{Phys. Dark Univ.} \textbf{44}, 101486 (2024).

\bibitem{2024PhRvD.109b3527A}
S.A. {Adil}, {\"O}.~{Akarsu}, E.~{Di Valentino}, R.C. {Nunes}, E.~{{\"O}z{\"u}lker}, A.A. {Sen} and E.~{Specogna}, \emph{Phys. Rev. D} \textbf{109}, 023527 (2024).

\bibitem{Nojiri:2022ski}
S.~Nojiri, S.~D. Odintsov and V.~K. Oikonomou, \emph{Nucl. Phys. B} \textbf{980}, 115850 (2022).

\bibitem{Odintsov:2020qzd}
Sergei~D. Odintsov, Diego S\'aez-Chill\'on~G\'omez and German~S. Sharov, \emph{Nucl. Phys. B} \textbf{966}, 115377 (2021).

\bibitem{2016RPPh...79i6901W}
B.~{Wang}, E.~{Abdalla}, F.~{Atrio-Barandela} and D.~{Pav{\'o}n}, \emph{Rep. Prog. Phys.} \textbf{79}, 096901 (2016).

\bibitem{naidoo2024PhRvD}
Krishna {Naidoo}, Mariana {Jaber}, Wojciech~A. {Hellwing} and Maciej {Bilicki}, \emph{Phys. Rev. D} \textbf{109}, 083511 (2024).

\bibitem{DAINOTTI202430}
M.G. Dainotti, G.~Bargiacchi, M.~Bogdan, S.~Capozziello and S.~Nagataki, \emph{Journal of High Energy Astrophysics} \textbf{41}, 30 (2024).

\bibitem{Dainotti2023ApJ...951...63D}
Maria~Giovanna {Dainotti}, Giada {Bargiacchi}, Malgorzata {Bogdan}, Aleksander~Lukasz {Lenart}, Kazunari {Iwasaki}, Salvatore {Capozziello}, Bing {Zhang} and Nissim {Fraija}, \emph{ApJ} \textbf{951}, 63 (2023).

\bibitem{bib:weinberg-2008}
Steven Weinberg, \emph{{Cosmology}} ({OUP Oxford}, {Oxford, England}) (2008).

\bibitem{DiValentino:2021izs}
Eleonora Di~Valentino, Olga Mena, Supriya Pan, Luca Visinelli, Weiqiang Yang, Alessandro Melchiorri, David~F. Mota, Adam~G. Riess and Joseph Silk, \emph{Class. Quant. Grav.} \textbf{38}, 153001 (2021).

\bibitem{apj-powerlaw}
Maria~Giovanna Dainotti, Biagio De~Simone, Tiziano Schiavone, Giovanni Montani, Enrico Rinaldi and Gaetano Lambiase, \emph{ApJ} \textbf{912}, 150 (2021).

\bibitem{galaxies10010024}
Maria~Giovanna Dainotti, Biagio De~Simone, Tiziano Schiavone, Giovanni Montani, Enrico Rinaldi, Gaetano Lambiase, Malgorzata Bogdan and Sahil Ugale, \emph{Galaxies} \textbf{10} (2022).

\bibitem{2024MNRAS.530.5091X}
Bing {Xu}, Jiancheng {Xu}, Kaituo {Zhang}, Xiangyun {Fu} and Qihong {Huang}, \emph{Mon. Not. RAS} \textbf{530}, 5091 (2024).

\bibitem{Krishnan:2020vaf}
C.~Krishnan, E.~\'O. Colg\'ain, M.~M. Sheikh-Jabbari and Tao Yang, \emph{Phys. Rev. D} \textbf{103}, 103509 (2021).

\bibitem{kazantzidis}
L.~Kazantzidis and L.~Perivolaropoulos, \emph{Phys. Rev. D} \textbf{102}, 023520 (2020).

\bibitem{2021MNRAS.505.3866E}
George {Efstathiou}, \emph{Mon. Not. RAS} \textbf{505}, 3866 (2021).

\bibitem{giare2024arXiv240412779G}
W.~{Giar{\`e}}, \emph{Phys. Rev. D} \textbf{109}, 123545 (2024).

\bibitem{2023Univ....9..393V}
Sunny {Vagnozzi}, \emph{Universe} \textbf{9}, 393 (2023).

\bibitem{2020PhRvD.102b3518V}
Sunny {Vagnozzi}, \emph{Phys. Rev. D} \textbf{102}, 023518 (2020).

\bibitem{Sotiriou-Faraoni:2010}
Thomas~P. Sotiriou and Valerio Faraoni, \emph{Rev. Mod. Phys.} \textbf{82}, 451 (2010).

\bibitem{NOJIRI201159}
Shin’ichi Nojiri and Sergei~D. Odintsov, \emph{Physics Reports} \textbf{505}, 59 (2011).

\bibitem{2001CQGra..18..193M}
Giovanni {Montani}, \emph{Classical and Quantum Gravity} \textbf{18}, 193 (2001).

\bibitem{1992PhLA..162..223C}
M.~O. {Calv{\~a}o}, J.~A.~S. {Lima} and I.~{Waga}, \emph{Physics Letters A} \textbf{162}, 223 (1992).

\bibitem{erdem24a}
Recai {Erdem}, \emph{arXiv e-prints} arXiv:2402.16791 (2024).

\bibitem{2015PhRvD..91f3526N}
Rafael~C. {Nunes} and Diego {Pav{\'o}n}, \emph{Phys. Rev. D} \textbf{91}, 063526 (2015).

\bibitem{2016GReGr..48..107N}
Rafael~C. {Nunes}, \emph{General Relativity and Gravitation} \textbf{48}, 107 (2016).

\bibitem{2014JCAP...10..042L}
J.~A.~S. {Lima}, L.~L. {Graef}, D.~{Pav{\'o}n} and Spyros {Basilakos}, \emph{Journal of Cosmology and Astroparticle Physics,} \textbf{2014}, 042 (2014).

\bibitem{2020MNRAS.496L..91E}
George {Efstathiou} and Steven {Gratton}, \emph{Mon. Not. RAS} \textbf{496}, L91 (2020).

\bibitem{birrelldavies}
N.D. {Birrell} and P.C.W. {Davies}, \emph{Quantum Fields in Curved Space} (Cambridge University Press) (1982).

\bibitem{WALDbook}
R.M. {Wald}, \emph{Quantum Field Theory in Curved Spacetime and Black Hole Thermodynamics} (The University of Chicago Press) (1994).

\bibitem{Olmo:2005hc}
Gonzalo~J. Olmo, \emph{Phys. Rev. D} \textbf{72}, 083505 (2005).

\bibitem{Brax:2008hh}
Philippe Brax, Carsten van~de Bruck, Anne-Christine Davis and Douglas~J. Shaw, \emph{Phys. Rev. D} \textbf{78}, 104021 (2008).

\bibitem{2017MNRAS.470.2617A}
Shadab {Alam}, Metin {Ata}, Stephen {Bailey}, Florian {Beutler}, Dmitry {Bizyaev}, Jonathan~A. {Blazek}, Adam~S. {Bolton}, Joel~R. {Brownstein}, Angela {Burden}, Chia-Hsun {Chuang}, Johan {Comparat}, Antonio~J. {Cuesta}, Kyle~S. {Dawson}, Daniel~J. {Eisenstein}, Stephanie {Escoffier}, H{\'e}ctor {Gil-Mar{\'\i}n}, Jan~Niklas {Grieb}, Nick {Hand}, Shirley {Ho}, Karen {Kinemuchi}, David {Kirkby}, Francisco {Kitaura}, Elena {Malanushenko}, Viktor {Malanushenko}, Claudia {Maraston}, Cameron~K. {McBride}, Robert~C. {Nichol}, Matthew~D. {Olmstead}, Daniel {Oravetz}, Nikhil {Padmanabhan}, Nathalie {Palanque-Delabrouille}, Kaike {Pan}, Marcos {Pellejero-Ibanez}, Will~J. {Percival}, Patrick {Petitjean}, Francisco {Prada}, Adrian~M. {Price-Whelan}, Beth~A. {Reid}, Sergio~A. {Rodr{\'\i}guez-Torres}, Natalie~A. {Roe}, Ashley~J. {Ross}, Nicholas~P. {Ross}, Graziano {Rossi}, Jose~Alberto {Rubi{\~n}o-Mart{\'\i}n}, Shun {Saito}, Salvador {Salazar-Albornoz}, Lado {Samushia}, Ariel~G. {S{\'a}nchez}, Siddharth {Satpathy},
  David~J. {Schlegel}, Donald~P. {Schneider}, Claudia~G. {Sc{\'o}ccola}, Hee-Jong {Seo}, Erin~S. {Sheldon}, Audrey {Simmons}, An{\v{z}}e {Slosar}, Michael~A. {Strauss}, Molly E.~C. {Swanson}, Daniel {Thomas}, Jeremy~L. {Tinker}, Rita {Tojeiro}, Mariana~Vargas {Maga{\~n}a}, Jose~Alberto {Vazquez}, Licia {Verde}, David~A. {Wake}, Yuting {Wang}, David~H. {Weinberg}, Martin {White}, W.~Michael {Wood-Vasey}, Christophe {Y{\`e}che}, Idit {Zehavi}, Zhongxu {Zhai} and Gong-Bo {Zhao}, \emph{Mon. Not. RAS} \textbf{470}, 2617 (2017).

\bibitem{2019A&A...629A..86B}
Michael {Blomqvist}, H{\'e}lion {du Mas des Bourboux}, Nicol{\'a}s~G. {Busca}, Victoria {de Sainte Agathe}, James {Rich}, Christophe {Balland}, Julian~E. {Bautista}, Kyle {Dawson}, Andreu {Font-Ribera}, Julien {Guy}, Jean-Marc {Le Goff}, Nathalie {Palanque-Delabrouille}, Will~J. {Percival}, Ignasi {P{\'e}rez-R{\`a}fols}, Matthew~M. {Pieri}, Donald~P. {Schneider}, An{\v{z}}e {Slosar} and Christophe {Y{\`e}che}, \emph{A \& A} \textbf{629}, A86 (2019).

\bibitem{2019A&A...629A..85D}
Victoria {de Sainte Agathe}, Christophe {Balland}, H{\'e}lion {du Mas des Bourboux}, Nicol{\'a}s~G. {Busca}, Michael {Blomqvist}, Julien {Guy}, James {Rich}, Andreu {Font-Ribera}, Matthew~M. {Pieri}, Julian~E. {Bautista}, Kyle {Dawson}, Jean-Marc {Le Goff}, Axel {de la Macorra}, Nathalie {Palanque-Delabrouille}, Will~J. {Percival}, Ignasi {P{\'e}rez-R{\`a}fols}, Donald~P. {Schneider}, An{\v{z}}e {Slosar} and Christophe {Y{\`e}che}, \emph{A \& A} \textbf{629}, A85 (2019).

\bibitem{2021MNRAS.500.1201H}
Jiamin {Hou}, Ariel~G. {S{\'a}nchez}, Ashley~J. {Ross}, Alex {Smith}, Richard {Neveux}, Julian {Bautista}, Etienne {Burtin}, Cheng {Zhao}, Rom{\'a}n {Scoccimarro}, Kyle~S. {Dawson}, Arnaud {de Mattia}, Axel {de la Macorra}, H{\'e}lion {du Mas des Bourboux}, Daniel~J. {Eisenstein}, H{\'e}ctor {Gil-Mar{\'\i}n}, Brad~W. {Lyke}, Faizan~G. {Mohammad}, Eva-Maria {Mueller}, Will~J. {Percival}, Graziano {Rossi}, Mariana {Vargas Maga{\~n}a}, Pauline {Zarrouk}, Gong-Bo {Zhao}, Jonathan {Brinkmann}, Joel~R. {Brownstein}, Chia-Hsun {Chuang}, Adam~D. {Myers}, Jeffrey~A. {Newman}, Donald~P. {Schneider} and M.~{Vivek}, \emph{Mon. Not. RAS} \textbf{500}, 1201 (2021).

\bibitem{2017PhR...711....1A}
Kevork~N. {Abazajian}, \emph{Phys. Rept.} \textbf{711}, 1 (2017).

\end{thebibliography}

\end{document}